# Single-Shot Electron Radiography

# Using a Laser-Plasma Accelerator

G. Bruhaug[1*], M. S. Freeman[2], H. G. Rinderknecht[1], L. P. Neukirch[2], C. H. Wilde[2], F. E Merrill[2], J. R. Rygg[1], M. S. Wei[1], G. W. Collins[1], and J. L. Shaw[1]

[1]Laboratory for Laser Energetics, University of Rochester, Rochester, NY 14623-1299, USA

[2]Los Alamos National Laboratory, Los Alamos, NM 87545, USA

**Abstract**

Contact and projection electron radiography of static targets was demonstrated using a laser-plasma accelerator driven by a kilojoule, picosecond-class laser as a source of relativistic electrons with an average energy of 20 MeV. Objects with areal densities as high as 7.7 g/cm$^2$ were probed in materials ranging from plastic to tungsten, and radiographs with resolution as good as 90 $\mu$m were produced. The effects of electric fields produced by the laser ablation of the radiography objects were observed and are well described by an analytic expression relating imaging magnification change to electric-field strength.

**Introduction**

Understanding high-energy-density (HED) plasmas, inertial confinement fusion (ICF) implosions, and laser–plasma interactions (LPI's) at large research facilities such as the OMEGA laser at the Laboratory for Laser Energetics (LLE), Laser Mégajoule at Commissariat à l'énergie atomique, the Z-machine at Sandia National Laboratories, and the National Ignition Facility at Lawrence

Livermore National Laboratory is important for mitigating the factors prohibiting ignition in the search for a sustainable fusion energy source [1], as well as to better understand other fundamental and radiation-driven physics. It is extremely difficult to characterize these events, which occur on a very small spatial scale (millimeter to micrometer) and very fast time scale (microsecond to picosecond), requiring a variety of diagnostic techniques that are constantly evolving.

To investigate the physical structure of compressed targets, laser-generated x-ray [2,3] or proton radiography [4–7] is typically used, with protons providing the extra feature of electromagnetic-field sensitivity. Although x-ray and proton probes are the standard laser-generated diagnostic, there is another laser-generated probe that has seen little use, namely, relativistic electrons. Small-scale HED research facilities have performed electron radiography of ultrafast laser–plasma interactions [7–12], but this capability has never before been extended to kilojoule- or megajoule-class laser facilities. The work presented in this manuscript builds upon previous electron radiography (eRad) work using radio-frequency (rf) linear accelerators [10,13−15] and small-scale lasers [7–9,11,12,16,17] and extends it to kilojoule-class facilities using the already available picosecond lasers for electron-beam generation via laser-plasma acceleration (LPA) [18,19].

Here, we report the first single-shot eRad images using an electron beam from a kilojoule-class LPA. Both contact and projection radiographs were obtained of static targets in materials ranging from plastic to tungsten, and resolutions as good as 90 $\mu$m were achieved. This work lays the foundation for future electron radiography of laser-driven targets at kilojoule- and megajoule-class facilities.

**Background**

Rf-powered linear accelerators generate monochromatic, low-emittance electron beams suitable for high-quality electron radiography [10,13–15]. Such systems, however, are rarely available at the same facilities as large HED drivers and cannot easily be installed for experiments because of cost and space constraints. Often these HED facilities have picosecond lasers available such as OMEGA EP, NIF-ARC (National Ignition Facility Advanced Radiograph Capability), PETAL, and Z-Petawatt lasers, which can be used for the efficient generation of relativistic electron beams via LPA techniques [18]. This method allows for electron beams to be generated for radiography without needing to add a rf linear accelerator to an HED facility. A laser-driven eRad system also possesses the temporal characteristics that could make for an ideal diagnostic of other picosecond-scale processes [7–9,11,12] for which linear accelerators do not typically provide equivalent instantaneous electron flux and may suffer signal-to-noise ratio issues.

Electron radiography provides a complementary probe to existing x-ray and proton radiography techniques. Laser-generated electrons are able to penetrate more material than laser-generated protons at the available energies, as shown in Fig. 1. For example, a laser-generated proton of 15 MeV will be fully stopped by ~2 mm of plastic at standard density and temperature, while a 15-MeV electron will require multiple centimeters of plastic to be fully stopped [20]. Far more electrons of similar energy or higher will also be generated for given laser conditions, providing a further advantage to laser-generated electrons over protons [5,18].

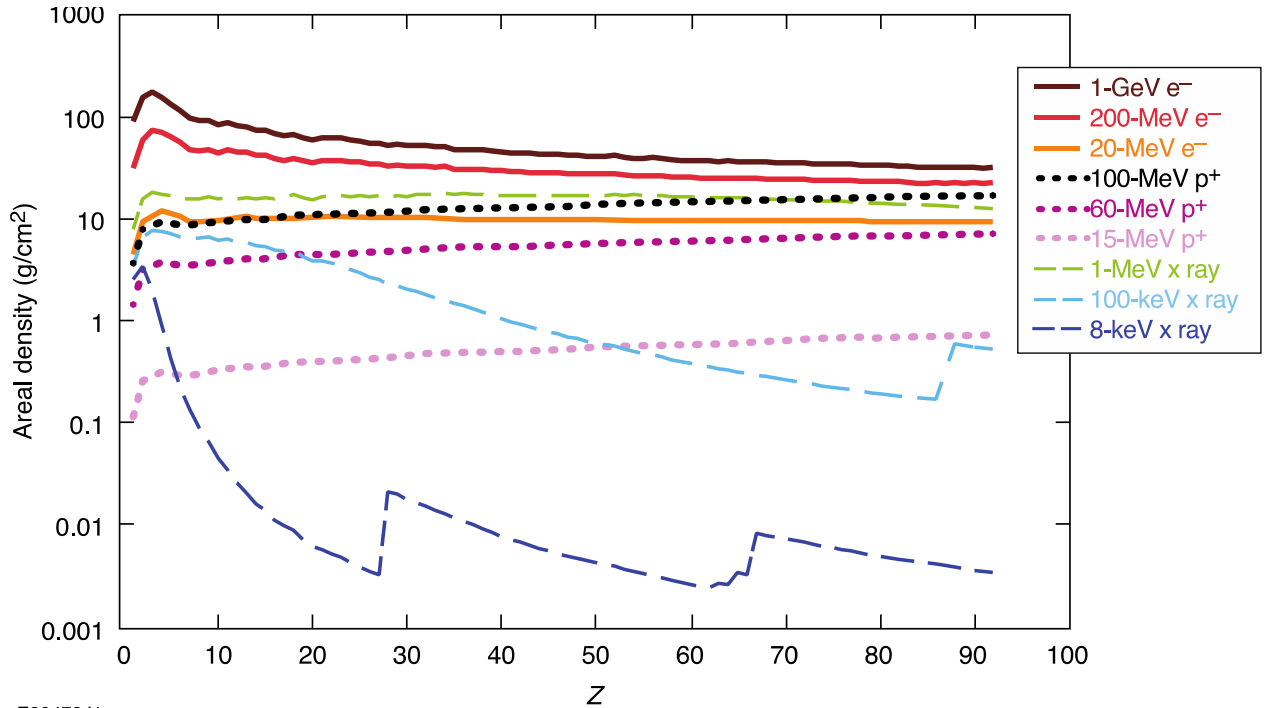

E30473J1

**Figure 1.** Areal density needed for a factor of $1/e$ reduction in particle flux versus atomic number ($Z$) for average, mid-scale, and high-probe-energy laser-driven electron, x-ray, and proton probes. Charged-particle ranges are determined via continuous slowing-down approximation, while x-ray ranges were determined with mass attenuation coefficients [20,21].

Typical mean/maximum probe energies generated on OMEGA-EP are 20/200 MeV for electrons, 15/60 MeV for protons and 8/100 keV for x-rays [2,3,18,22–24]. It can be seen in Fig. 1 that relativistic electrons are the most penetrating probes able to be generated with currently available lasers at HED facilities [2,3,18,22–24]. This penetration capability is crucial for radiographing targets at extreme densities, such as those used for ICF experiments [25–27]. Megavolt-scale x rays can have similar penetrative capabilities to that of relativistic electrons [3] but are typically generated by first generating relativistic electrons of a much higher energy. Thus, more electrons can be sent to the target for the same laser input power verses Megavolt-scale x rays, and the signal-to-noise ratio can potentially be increased for the same experimental parameters.

Relativistic electrons are more sensitive to deflection by electric and magnetic fields for a given energy than protons, but have a higher ratio of electric to magnetic field sensitivity. The higher ratio of the magnetic- versus electric-field deflection sensitivity makes laser- generated electrons an excellent complement to laser protons for radiography of electromagnetic fields. This can be seen by comparing the magnetic rigidity $B\rho \equiv B$ (d$x$/d$\theta$) (the resistance of a charged particle to deflection from a magnetic field) to the equivalent electric-field deflection resistance $E\rho \equiv E$ (d$x$/d$\theta$) ("electric rigidity"), where $B$ and $E$ are the magnetic and electric fields and d$x$ and d$\theta$ are change in position and angle, respectively [28–30]. Deriving in the limit of small deflections, the magnetic and electric rigidity are

$$B\rho = \frac{p}{q} \tag{1}$$

and

$$E\rho = {pc\beta}/{q} \tag{2}$$

The units for magnetic rigidity are Tesla meters per radians and the input is relativistic momentum $p$ and charge $q$ of the particle in question [28,30]. Electric rigidity depends on the particle momentum $p$, the particle mass $m$, the particle charge $q$, the velocity fraction of light speed $\beta$, and the speed of light $c$. The units of electric rigidity are volt-meter/meter per radian (in the present case, megavolt-meter/meter per radian is appropriate). The radian term is conventionally dropped when discussing magnetic rigidity [28,30] and will also be dropped for the remainder of the paper for electric rigidity as well. Figure 2 shows a comparison of magnetic and electric rigidity for both protons and relativistic electrons with notable electron and proton energies from various laser and rf sources [31] included. Energies up to 1 GeV are shown corresponding to today's high-performing electron LPA's [23,32] and large proton LINACs [31]. LPA-generated electron beams are regularly created with energies as low as a few MeV [33] and as high as ~8 GeV [18,23,34]. This capability for large variation in energy makes for a customizable radiography tool that can provide insight into a wide variety of targets at areal densities ranging from mg/cm$^2$ to many g/cm$^2$, integrated magnetic fields from $4.9 \times 10^{-3}$ to 28 T-m, and integrated electric fields from 0.9 to 1000 MV-m/m based on the above variation in electron energy.

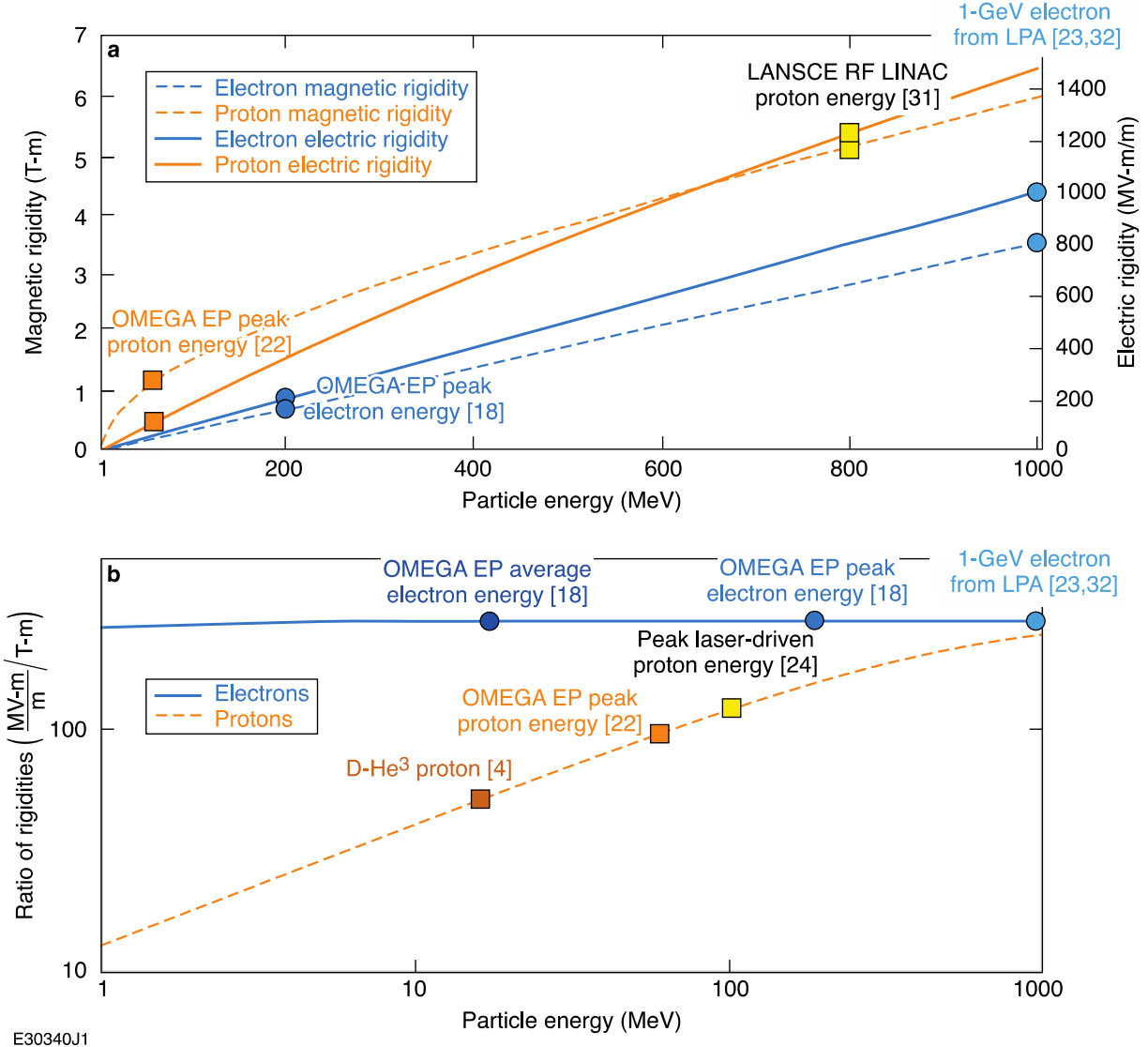

**Figure 2** (a) Magnetic and electric rigidity of protons and relativistic electrons up to 1 GeV. (b) The ratio of rigidities for selected laser-driven proton and relativistic electron sources, with a focus on OMEGA EP because of its common use as a proton source for HED and LPI experiments.

Relativistic electron radiography makes a great complement to proton radiography for laser-plasma experiments because of the much larger ratio of electric to magnetic rigidity compared to available protons at HED facilities [4–6, 22, 31]. The peak electron energies available [18, 23, 32] also have a higher electric rigidity than the peak proton energies available at HED facilities [4-7]. With eRad, suspected fields can be probed with an entirely different type of charged-particle probe, confirming or disproving theories about field type and strength.

The sensitivity of relativistic electrons to magnetic fields also provides for the option to use magnetic optics to improve resolution. Without the use of magnetic optics, charged-particle radiography is limited to the inherent resolution of the source size combined with the scatter-induced object blur and any imaging system resolution limits. Resolutions as low as 8.8 $\mu$m have been seen in previous eRad experiments with magnetic optics [14], and potentially ~0.06 $\mu$m with further refinement [13].

For radiography objects much smaller than the radiation length (the characteristic amount of material that a given charged particle can traverse before losing $1/e$ of its energy) [36], the resolution will be dominated by the size of the source of particles [13]. This limitation is also commonly seen in laser proton radiography experiments and provides resolutions comparable to the drive laser spot size of several $\mu$m [37]. If the target is thick enough to provide multiple scatters of the source particles, then the radiography resolution becomes dominated by scatter-based blurring. Optimal electron radiography occurs near $1/10^{\text{th}}$ of a radiation length, while ~1/2 of a radiation length ends up absorbing nearly all of the probing radiation [13]. The theoretical resolution of electron radiography can be modeled with the following analytic equations [35]:

$$\text{resolution} = \sqrt{x^2 + ip^2 + s^2} \tag{3}$$

$$x = \frac{d}{\gamma M} \sqrt{h16\pi NZ(Z+1) r_e^2 \ln\left(204 / Z^{0.33}\right) / \gamma^2 \beta^4} \tag{4}$$

$$ip = \frac{\Delta I}{M} \qquad (5)$$

and

$$s = \Delta S \frac{M-1}{M} \qquad (6)$$

Theoretical resolution is determined by adding the contributions of electron scattering ($x$), image plate (or other imaging system) resolution ($ip$) and source size ($s$) in quadrature in Eq. (3). Scatter-based blur is determined by Eq. (4), where $d$ is the drift distance from the radiography target to the imaging plane, $M$ is the magnification of the image, $h$ is the thickness of the target, $N$ is the number density of the material, $Z$ is the atomic number, $r_e$ is the classical electron radius, and $\gamma$ is the relativistic Lorentz factor. Equation 5 provides for imaging-system–induced blur with $\Delta I$ being the imaging system resolution in the image plane and $M$ is the magnification of the object as before. Source resolution is then determined with Eq. 6 with $\Delta S$ being the size of the source and $M$ is the magnification of the object.

These equations do not account for particle energy spread or bremsstrahlung-induced background, which have both been found to affect the resolution during eRad experiments [14]. The use of magnetic optics can mitigate the blur over a limited energy band by focusing the scattering particles onto a focal plane as well as eliminating blur associated with geometric and magnification effects. Particles outside of the focal energy range can be eliminated through the use of an energy collimator within a bending magnet and the lowest energy particles absorbed with filters, which also has the effect of bending the charged-particle focal plane outside of the line of sight of the unwanted bremsstrahlung radiation generated by the electron–target interaction [13].

**Experimental Setup**

The experiments were performed using the OMEGA EP laser, which has a central wavelength of 1054 nm and a pulse duration of 700±100 fs. The laser was spatially apodized from the typical $f/2$ geometry to $f/5$ to improve the focal quality and increase the Rayleigh length. Laser pulse energy was varied between 25 and 120 J and the $R_{80}$ spot size (the radius of the area containing 80% of the laser energy) varied in the range 13.9 to 16.2 $\mu$m. The laser was focused 500 $\mu$m inside a Mach-5 gas jet produced by a 6-mm nozzle with gas pressures ranging from 80 to 350 psi. This configuration has been found to produce a polychromatic, microcoulomb-class electron beam with a mean energy of 20±5 MeV with some extreme outliers at ±10 MeV [18]. Several measured samples of the electron energy spectra can been in Appendix A.

The experiment was performed in two different configurations: a contact eRad configuration and a projection eRad configuration. For the contact configuration (Fig. 3a), radiography test objects were placed in a detector stack consisting of 12.5-$\mu$m Al shielding, two MS image plates (IP's), and the radiography test objects (Fig. 3b), followed by an additional MS image plate. The detector pack was held at a distance of 56 cm from the laser focus in front of an electron–positron–proton spectrometer (EPPS) [38]. The image plates in front of the radiography test objects are used to measure the nonuniform transverse electron beam profile, which could then be subtracted from the radiograph to flatten the image for clarity [39]. The radiography test objects contained a series of steps ranging in thickness from 0.571±0.127 mm to 4.000±0.127 mm in 0.571±0.127-mm steps Each step thickness contained a series of holes with diameters ranging from 1.270±0.076 mm to 2.540±0.076 mm. The target design was chosen to provide a variety of features to measure the radiographic performance of the LPA electron beam in contact radiography geometry. Six different materials were radiographed: Cu (110 series), Sn, Al (1100 series), W (MT-17C), Ti, and stainless-steel 304L, covering a wide range of $Z$ numbers and densities.

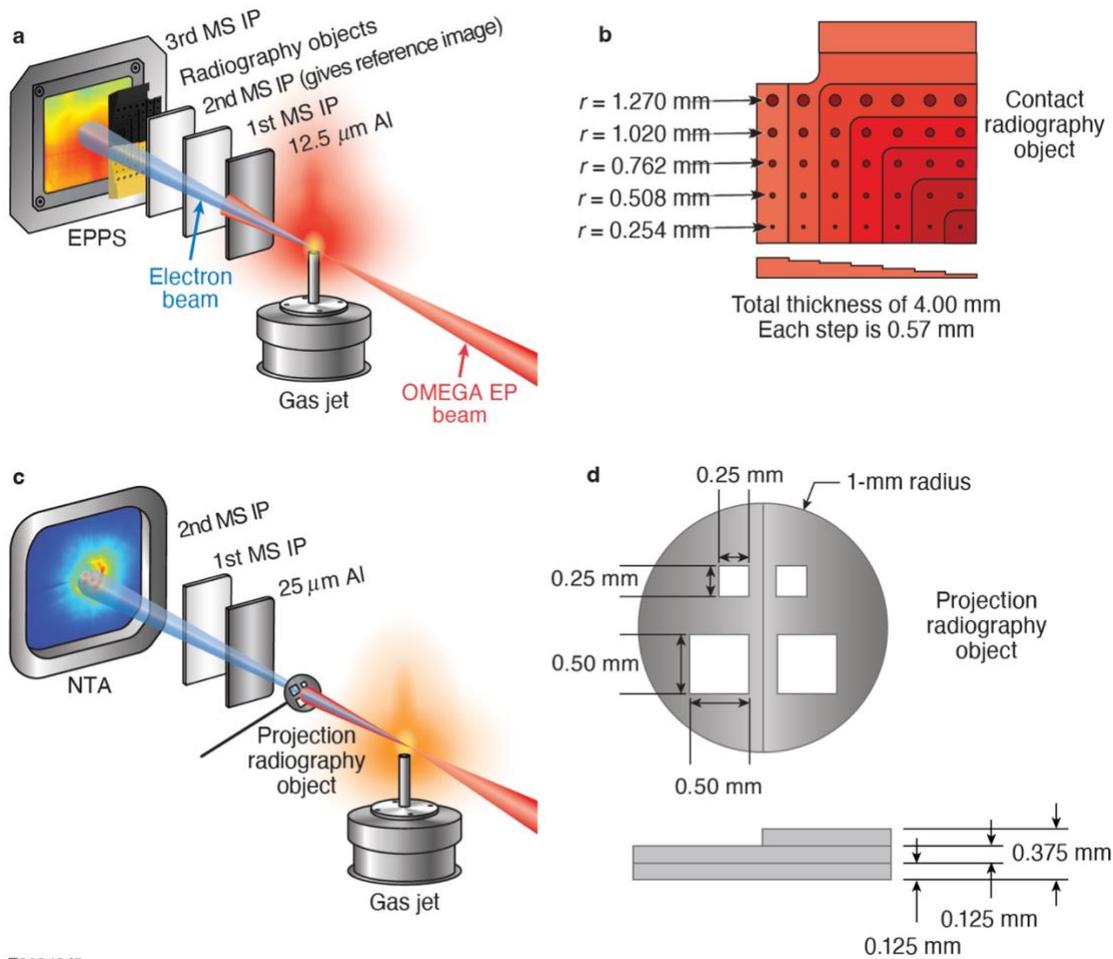

**Figure 3.** (a) Experimental setup for contact LPA eRad using radiography test objects (b) placed directly onto image plates and (c) projection LPA eRad using 2-mm-diam radiography test objects (d) offset from the image plates by distances ranging from 3.58 to 33.58 cm.

For the projection radiography experiments shown in Fig. 3c, radiography test objects (Fig. 3d) were placed 14.2 mm from laser focus and mounted on silicon carbide stalks. Note that the majority of the drive laser energy is transmitted through the gas jet and impacts the projection radiography object. It is estimated that ~20 to ~100 J of laser energy at intensities of ~$3 \times 10^{14}$ to ~$1 \times 10^{15}$ W/cm$^2$ impacted the radiography test objects with the laser beam head arriving ~45 fs before the head of the electron beam [40] and the remaining laser overlapping the electron beam. The laser then drives electric fields of the order of ~1 GV/m in the plasma sheath generated on the front face of the object. The electron beam was imaged via two stacked image plates wrapped in 25-$\mu$m aluminum foil placed in a near target arm (NTA) and varying in imaging distance from 3.58 to 33.58 cm from the radiography test object.

These radiography test objects were much smaller and thinner than the contact radiography test objects because of debris concerns for the laser optics. These smaller and thinner test objects are much more realistic stand-ins for future laser-driven HED targets and were made of the same W, Cu, Ti as above as well as solution-cast polystyrene to cover a wide range of target $Z$ and density options. The radiography objects had thicknesses ranging from 0.125±0.012 mm to 0.375±0.037 mm, areal densities ranging from 0.025 g/cm$^2$ to 0.713 g/cm$^2$ and hole sizes ranging from 0.25±0.03 mm to 0.5±0.03 mm (Fig. 3d). The radiation lengths, radiographed thicknesses, and optimal radiation lengths (~10% of a full radiation length) for all materials radiographed in these experiments are shown in Fig. 4.

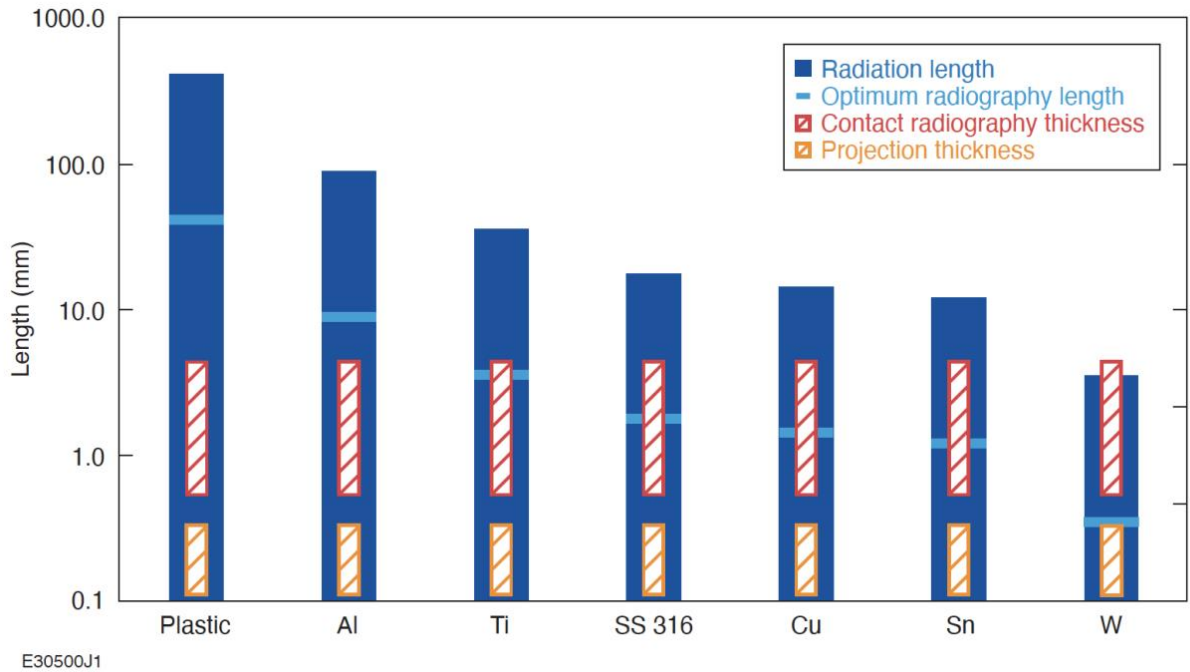

E30500J1

**Figure 4** Radiation length [Ref. 36] and optimal radiography length [Ref. 13] for all materials radiographed with the OMEGA EP LPA electron beam.

It can be seen that the lowest *Z* projection and contact radiograph objects are nowhere near a radiation length, or even a significant fraction of one, which will result in very little scattering. However, the W projection radiography objects are already effectively at an optimum thickness for scatter-dominated radiography and the W contact radiography object achieves a full radiation length at the thickest. Thus, little electron transmission can be expected through the thickest tungsten objects, while the thinnest polystyrene objects will be expected to barely scatter any electrons at all.

**Results and Discussion**

In both configurations, resolution is repeatedly measured by fitting an error function to a box out taken over edges and features of interest [39]. Total object sizes are then measured to check for magnification errors. Error in measurements is derived via repeated measurements of the same section and then the standard deviation of these measurements used.

Figure 5 shows that the resolution in the contact eRad configuration degraded with increasing thickness and $Z$ number of radiography target in rough agreement with the theoretical predictions, although there were outliers in the thinner tungsten sections. The poor fit of the theoretical prediction with the thinner tungsten sections can be explained via bremsstrahlung blurring, where the x-rays generated from the electrons interacting with the target decrease the signal-to-noise ratio and increase the effective source size due to their cone of emission. Mid- to high-$Z$ materials in this thickness range are considered ideal electron beam bremsstrahlung convertors for x-ray production for applications [41,42] and thus the blurring will be worse. Thicker sections will self-absorb the bremsstrahlung and trend toward the idealized predictions from Eqs. (3)–(6), as seen by the relatively good fits with the thickest tungsten sections. This is further supported by the much better theoretical fit with the aluminum sections, which had minimal bremsstrahlung production. Bremsstrahlung blurring has motivated the use of magnetic optics to separate out the electrons from the x-ray source in previous eRad experiments [13].

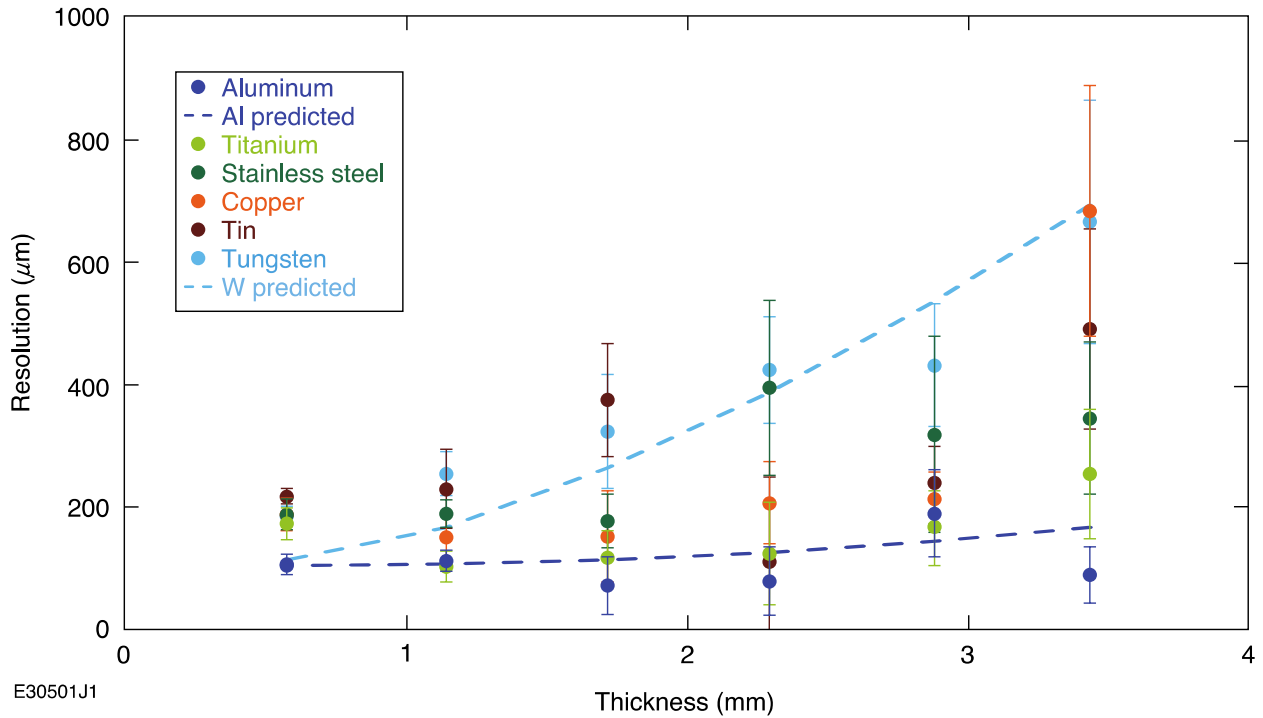

**Figure 5.** Resolution of contact radiography test object versus target thickness. Theoretical predictions based on Eqs. (3)–(6) for the tungsten and aluminum objects are included to guide the eye at the extremes of contact radiography test object $Z$ numbers.

The drift distance was taken to be the radiography object thickness at the point of measurement with a magnification of 1 using Eqs. (3)–(6) for calculating theoretical resolution. Image plate pixel size is taken to be ~100 $\mu$m [Ref. 43] and the source size is taken to be that of the laser, which ranged from 13.9 to 16.2 $\mu$m.

The effect of radiography object $Z$ number on projection radiography was tested, as seen in Fig. 6. All radiography objects were imaged with the NTA 6.58 cm from the object and 8 cm from the laser focal point. The electron beam is assumed to be born from the exit of the gas jet and is treated as a point source equal in size to the laser focal spot. For theoretical calculations of resolution, source size and pixel size were the same as above, but magnification was 5.3 and object thickness taken to be the average of 0.3125 mm for the radiography object. The resolution was found to be insensitive to the radiography test object $Z$ number.

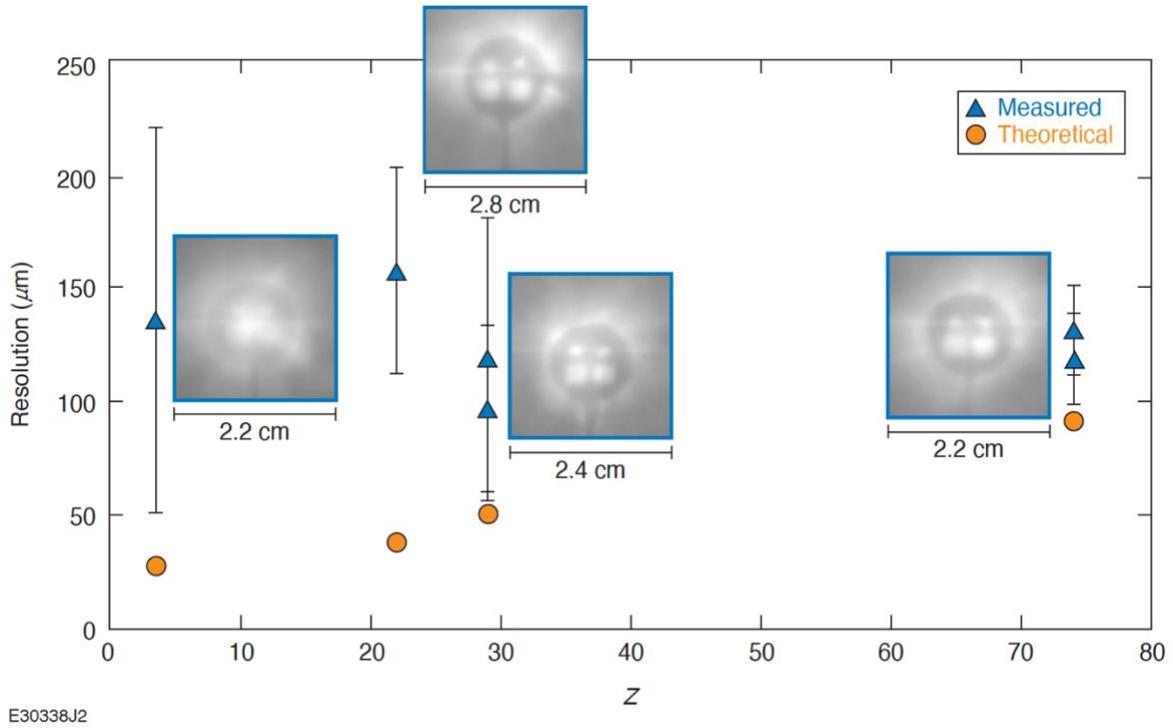

**Figure 6.** Resolution in the object plane versus atomic number ($Z$) of the target material for the projection configuration when the image plate was 8 cm from the location of the best laser focus. Theoretical predictions are calculated using Eq. (3). Each data point has the radiograph recorded on the image plate next to it.

This response is promising for future use as a radiography platform for laser-driven targets but does not match theoretical expectations from Eqs. (3)–(6), which predict a roughly linear relationship with radiography test object $Z$ for the same object thickness (Fig. 6). It should be noted that the objects vary over an order of magnitude in areal density as the $Z$ number increases. This flat response with $Z$ could be caused by electron-plasma–induced blur from the drive laser generating a surface plasma on the front face of the radiography test object or from the above-mentioned bremsstrahlung blurring dominating over the minimal scatter of the electrons that would happen in such low density, low-Z objects.

To test the radiographic capabilities of the projection configuration at different imaging distances, repeat radiographs of a tungsten radiography test object were performed with the image plate pack placed at 3.58, 6.58, 22.08 and 33.58 cm from the radiography test object (Fig. 7). Theoretical resolution was calculated using the same source size, pixel size and object thickness as for Fig. 6, but now with varying magnifications. The resolution held roughly constant with imaging distance, indicating that resolution is limited by scattering rather than by source size and that gains in resolution can only be had with the inclusion of magnetic optics. Although the exact resolutions were not well predicted by the theoretical predictions, the general trend was captured. As imaging distance increases, the magnification increases, which improves resolution. This improvement is countered, however, by the increased distances allowing the scattered electrons to undergo further displacement and thus increasing the imaging blur, degrading the resolution of the image. This is the primary motivation for the addition of magnetic optics in charged-particle radiography schemes as the large angle scattered particles are removed from the image and the small angle scatters are refocused back onto the imaging plane [13]. Bremsstrahlung blurring can also be avoided by imaging the beam outside of the direct line of sight of the radiography target using dipole magnets. In this case, no such optics were used and, as such, the resolution remains limited to 90 $\mu$m in the best radiographs with W radiography objects and the NTA placed 3.58 cm from the radiography object as seen in Fig. 7.

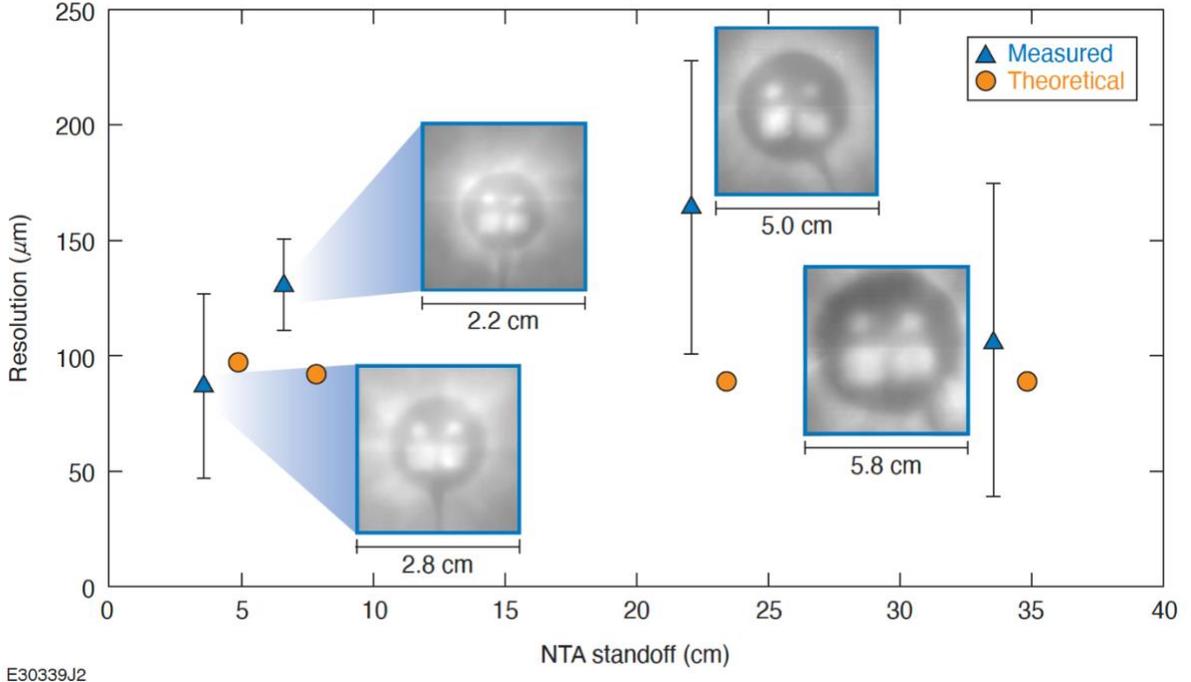

E30339J2

**Figure 7.** Resolution in the object plane versus imaging distance for tungsten projection radiography objects. Theoretical predictions are calculated using Eq. (3). Each data point has the radiograph recorded on the image plate next to it.

An additional phenomenon of interest was seen with projection radiography during analysis. The measured magnification of the image is persistently ~1.5× smaller than would be expected from the distance between the source, target, and image plate. We conjecture that electrostatic focusing from laser-plasma generated fields on the front face of the radiography object is the cause of this discrepancy. Laser-generated fields of the order of ~1 GV/m have been reported at similar laser intensities as are generated on the front face of the radiography objects [4,44] and G4 Beamline simulations [45] indicate similar changes in image magnification at similar electric-field strengths. Using the concept of electrostatic rigidity [Eq. (2)], we can derive an analytic estimate for the electric field corresponding to the observed deflection, which is given by Eq. (7).

Full details of this derivation are included in Appendix B and representative simulations results in Appendix C.

$$E\left(\frac{V}{m}\right) = \left(\frac{y}{\sqrt{y^2+r^2}}\right)\left(\frac{R'-R}{x}\right)\frac{pc\beta}{\delta xq} \tag{7}$$

Here, $R$ and and $R'$ are the expected and measured size of the radiography target, respectively, while $r$ is the original size of the object being analyzed. Object radius is $r$, while $y$ and $x$ are the distance from the source to the image plane and target to the image plane, respectively. The variables $p$, $q$, and $\beta$ are the same relativistic and particle specific terms from Eq. (2) and $\delta x$ is the assumed field length. Note that changes in the assumed field length drastically change the predicted electric field. This equation does not account for the gradients actually expected in the plasma-generated electric fields [44] and an average field is assumed for a given distance. Assuming a weighted average electron energy of 20$\pm$5 MeV[18], a field length of 0.3 mm (an average length of the target) for $\delta x$, and using the experimental parameters for each shot taken, an estimate of the required electric field can be made and was found to average ~3 GV/m. The estimated electric fields were plotted versus laser energy in Fig. 8.

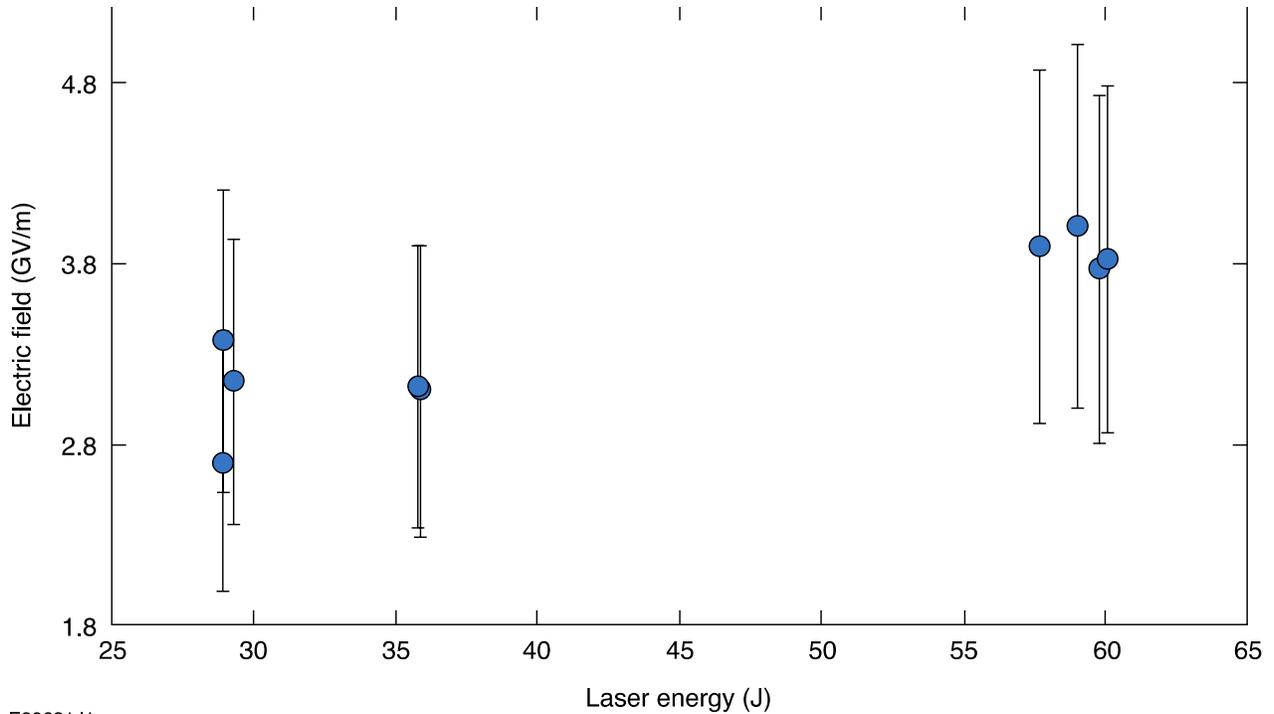

**Figure 8** Estimated electric field to cause magnification change versus laser-drive energy used in the experiment.

Estimated electric field increased roughly linearly with laser energy on target within a specific intensity range, as expected by previous studies on the topic [44]. These results cover a wide range of materials, which can help to explain the variations in measured electric field for a given laser energy. The analytic method derived here can be used to estimate large-scale electric fields in future laser-driven experiments and is applicable to all forms of charged-particle radiography.

**Conclusion and Future Work**

An eRad platform based on an LPA driven by a kilojoule, picosecond-class laser was tested on a wide variety of targets and imaging configurations. Resolutions as good as 90 $\mu$m were seen for projection radiography with a magnification of 3.61. Imaging resolution degraded with thickness of the target material and with $Z$ number of the material as expected. Increased magnification did not improve resolution, indicating that resolution is limited by scattering rather than by source size. This motivates the addition of magnetic optics and collimators on future experiments to mitigate angular blur and improve resolution. Electric fields from laser irradiation of the radiography target were also measured, and an analytic equation derived to estimate electric fields from the change in magnification of the image. Future experiments will radiograph a wide variety of laser-driven targets to measure electric- and magnetic-field generation in laser–solid interactions and strongly shocked materials for ICF and HED applications.

## Acknowledgement

This material is based upon work supported by the Department of Energy National Nuclear Security Administration under Award Number DE-NA0003856 and under Award Number 89233218CNA000001, the U.S. Department of Energy under Awards DE-SC00215057, the University of Rochester, and the New York State Energy Research and Development Authority. Partial funding for G. W. Collins and J. R. Rygg was from US Department of Energy, Office of Science, Fusion Energy Sciences under award number DE-SC0020340.

by the U.S. Government or any agency thereof. The views and opinions of authors expressed herein do not necessarily state or reflect those of the U.S. Government or any agency thereof.

**Data Availability**

The data that support the plots within this paper and other finding of this study are available from the corresponding author upon reasonable request.

**References**

1.     Edwards, M. J. *et al.* Progress towards ignition on the National Ignition Facility. *Phys. Plasmas* **20**, 070501 (2013).

2.     Dewald, E. L. *et al.* X-ray streaked refraction enhanced radiography for inferring inflight density gradients in ICF capsule implosions. *Rev. Sci. Instrum.* **89**, 10G108 (2018).

3.     Courtois, C. *et al.* High-resolution multi-MeV x-ray radiography using relativistic laser-solid interaction. *Phys. Plasmas* **18**, 023101 (2011).

4.     Rygg, J. R. *et al.* Proton radiography of inertial fusion implosions. *Science* **319**, 1223−1225 (2008).

5.     Zylstra, A. B. *et al.* Using high-intensity laser-generated protons to radiograph directly driven implosions. *Rev. Sci. Instrum.* **83**, 013511 (2012).

6.     Li, C. K. *et al.* Proton radiography of dynamic electric and magnetic fields in laser-produced high-energy-density plasmas. *Phys. Plasmas* **16**, 056304 (2009).

7.     Wan, Y. *et al.* Direct observation of relativistic broken plasma waves. *Nat. Phys.* **18**, 1186−1190 (2022).

8.     Schumaker, W. *et al.* Ultrafast electron radiography of magnetic fields in high-intensity laser-solid interactions. *Phys. Rev. Lett.* **110**, 015003 (2013).

9.      Raj, G. *et al.* Probing ultrafast magnetic-field generation by current filamentation instability in femtosecond relativistic laser-matter interactions. *Phys. Rev. Research* **2**, 023123 (2020).

10.     Zhang, C. *et al.* Measurements of the growth and saturation of electron Weibel instability in optical-field ionized plasmas. *Phys. Rev. Lett.* **125**, 255001 (2020).

11.     Zhou, Z. *et al.* Visualizing the melting processes in ultrashort intense laser triggered gold mesh with high energy electron radiography. *Matter Radiat. Extremes* **4**, 065402 (2019).

12.     Zhang, C. J *et al.* Femtosecond probing of plasma wakefields and observation of the plasma wake reversal using a relativistic electron bunch. *Phys. Rev. Lett.* **119**, 064801 (2017).

13.     Merrill, F. E. Imaging with penetrating radiation for the study of small dynamic physical processes. *Laser Part. Beams* **33**, 425–431 (2015).

14.     Merrill, F. E. *et al.* Demonstration of transmission high energy electron microscopy. *Appl. Phys. Lett.* **112**, 144103 (2018).

15.     Merrill, F. *et al.* Electron radiography. *Nucl. Instrum. Methods Phys. Res. B* **261**, 382–386 (2007).

16.     Hazra, D., Mishra, S., Moorti, A. & Chakera, J. A. Electron radiography with different beam parameters using laser plasma accelerator. *Phys. Rev. Accel. Beams* **22**, 074701 (2019).

17.     Xiao, C. *et al.* Direct imaging with hundreds of MeV electron bunches from laser wakefield acceleration. *Phys. Stat. Sol. (A)* **218**, 2000371 (2021).

18.     Shaw, J. L. *et al.* Microcoulomb ($0.7 \pm 0.4/0.2 \ \mu c$) laser plasma accelerator on OMEGA EP. *Sci. Rep.* **11**, 7498 (2021).

19. Albert, F. *et al.* Betatron x-ray radiation in the self-modulated laser wakefield acceleration regime: Prospects for a novel probe at large scale laser facilities. *Nucl. Fusion* **59**, 032003 (2018).

20. Berger, M. J., Coursey, J. S., Zucker, M. A. & Chang, J. (2017), *ESTAR*, *PSTAR*, and *ASTAR*: Computer Programs for Calculating Stopping-Power and Range Tables for Electrons, Protons, and Helium Ions (Ver. 2.0.1). [Online] Available: https://www.nist.gov/pml/stopping-power-range-tables-electrons-protons-and-helium-ions [17 August 2018]. National Institute of Standards and Technology, Gaithersburg, MD, Accessed 1 April 2022, https://dx.doi.org/10.18434/T4NC7P.

21. Berger, M. J. *et al.* Xcom: Photon cross sections database. NIST Standard Reference Database 8 (XGAM) (ver. 3.1), Accessed 1 April 2022, https://dx.doi.org/10.18434/T48G6X.

22. Flippo, K. *et al.* OMEGA EP, laser scalings and the 60 MeV barrier: First observatios of ion acceleration performance in the 10 picosecond kilojoule short-pulse regime. *J. Phys.: Conf. Ser.* **244**, 022033 (2010).

23. Gonsalves, A. J. *et al.* Laser-heated capillary discharge plasma waveguides for electron acceleration to 8 GeV. *Phys. Plasmas* **27**, 053102 (2020).

24. Higginson, A. *et al.* Near-100 MeV protons via a laser-driven transparency-enhanced hybrid acceleration scheme. *Nat. Commun.* **9**, 724 (2018).

25. Clark, D. S. *et al.* Three-dimensional simulations of low foot and high foot implosion experiments on the National Ignition Facility. *Phys. Plasmas* **23**, 056302 (2016).

26. Craxton, R. S. *et al.* Direct-drive inertial confinement fusion: A review. *Phys. Plasmas* **22**, 110501 (2015).

27. McBride, R. D. *et al.* A primer on pulsed power and linear transformer drivers for high energy density physics applications. *IEEE Trans. Plasma Sci.* **46**, 3928–3967 (2018).

28. Lee, S. Y. *Accelerator physics*. 2nd ed. (World Scientific Publishing Company, New Jersey, 2004).

29. Kugland, N. L., Ryutov, D. D., Plechaty, C., Ross, J. S. & Park, H.-S. Invited article: Relation between electric and magnetic field structures and their proton-beam images. *Rev. Sci. Instrum.* **83**, 101301 (2012).

30. Krinsky, S. Particle accelerator physics. *Synchrotron Radiat. News* **7**, 39A (1994).

31. King, N. S. P. *et al.* An 800-MeV proton radiography facility for dynamic experiments. *Nucl. Instrum. Methods Phys. Rev. A* **424**, 84–91 (1999).

32. Kim, H. T. *et al.* Stable multi-GeV electron accelerator driven by waveform-controlled PW laser pulses. *Sci. Rep.* **7**, 10203 (2017).

33. Bussolino, G. C. *et al.* Electron radiography using a table-top laser-cluster plasma accelerator. *J. Phys. D: Appl. Phys.* **46**, 245501 (2013).

34. Esarey, E., Schroeder, C. B. & Leemans, W. P. Physics of laser-driven plasma-based electron accelerators. *Rev. Mod. Phys.* **81**, 1229–1285 (2009).

35. Nassiri, A. Stopping power and scattering angle calculations of charged particle beams through thin foils. Argonne National Laboratory, Urbana, IL, Report LA-165 (1991).

36. Workman, R. L. *et al.* Review of particle physics. *Prog. Theor. Exp. Phys.* **2022**, 083C01 (2022).

37. Campbell, P. T. *et al.* Proton beam emittance growth in multipicosecond laser-solid interactions. *New J. Phys.* **21**, 103021 (2019).

38.    Ayers, S. L. Electron positron proton spectrometer for use at laboratory for laser energetics. Lawrence Livermore National Laboratory, Livermore, CA, Report LLNL-TR-427769 (2010).

39.    Bruhaug, G. M. Laser-plasma-accelerator-driven electron radiography on the OMEGA EP laser. *presented at the North American Particle Accelerator Conference, Albuquerque, NM, 7–12 August 2022* (2022).

40.    N. Lemos *et al.*, "Self-modulated laser wakefield accelerators as x-ray sources," *Plasma Phys. Control. Fusion*, vol. 58, no. 3, 2016, doi: 10.1088/0741-3335/58/3/034018.

41.    Tsechanski, A., Fedorchenko, D., Starovoitova, V. & Galperin, A. Converter optimization for photonuclear production of Mo-99. *Nucl. Instrum. Methods Phys. Res. B* **461**, 118–123 (2019).

42.    Halbleib, J. A., Lockwood, G. J. & Miller, G. H. Optimization of bremsstrahlung energy deposition. *IEEE Trans. Nucl. Sci.* **23**, 1881–1885 (1976).

43.    Fiksel, G., Marshall, F. J., Mileham, C. & Stoeckl, C. Note: Spatial resolution on fuji bas-tr and bas-sr imaging plates. *Rev. Sci. Instrum.* **83**, 086103 (2012).

44.    Dubois, J. L. *et al.* Target charging in short-pulse-laser–plasma experiments. *Phys. Rev. E* **89**, 013102 (2014).

45.    Roberts, T. G4 beamline 3.06. Accessed 24 June 2022, https://www.muonsinc.com/Website1/G4beamline.

## Appendix A

The electron energy spectrum has been extensively measured in previous experiments with this platform and several representative spectra are shown in Fig 1A. This sample has two "typical" results (samples 1 and 2) and one low energy outlier (sample 3) to show the potential variation.

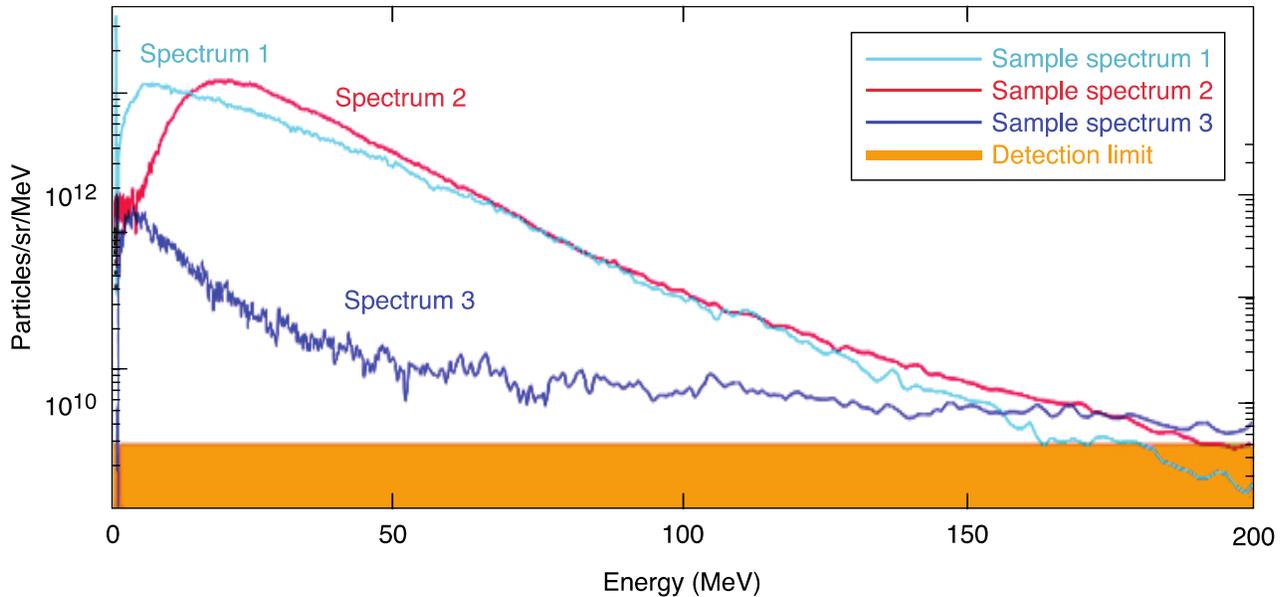

**Figure 1A.** Several sample electron energy spectra from the OMEGA EP generated electron beam

The weighted average energy is 20 MeV with an average energy variation of $\pm 5$ MeV, but there are outliers that exceed this range on some shots (such as sample 3). The energy spread and variation in weighted average electron energy makes the use of this beam for field measurements challenging and motivates further upgrades to the system to generate a more monochromatic beam with greater reliability in generating the same peak energy.

## Appendix B

To derive Eq. 7, one first starts with Eq. 2 for electric rigidity.

$$E\rho = \frac{pc\beta}{q} \qquad (1A)$$

Using the mean-particle momentum of the beam, a radius of curvature from a given electric-field relationship is established. We can then solve for electric field and achieve the following:

$$E = \frac{pc\beta}{q\rho} \qquad (2A)$$

We then treat the change in momentum and curvature as small changes by assuming the radiography object is acting as a very thin lens compared to the distances involved. The term $\delta\rho$ will also be changed to $\delta x$ to eliminate confusion with the term $\delta p$.

$$E \approx \frac{\delta p}{\delta x}\frac{c\beta}{q} \qquad (3A)$$

Fig. 2A shows the basic geometry of the derivation, where $y$ is the distance from source to radiography object, $x$ is the distance from the radiography object to the imaging plane, $r$ is the radius of the radiography object, $R$ is the expected radius of the image, and $R'$ is the actual radius of the image seen.

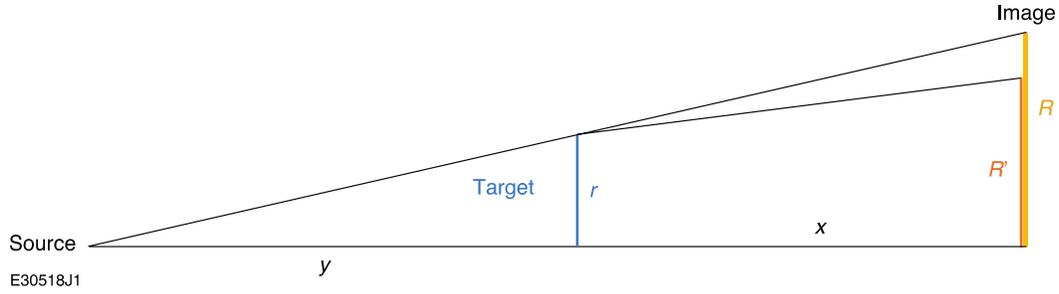

**Figure 2A.** Outline of radiography geometry for electric-field alterations of the image

Using Fig 2A the radius of curvature of the beam can be derived by first determining the angle between the unaffected trajectory (R) and the new trajectory (R') utilizing the trapezoid created by $r$, $x$ and $R'$.

Utilizing this geometry and the thin lens approximation and the geometry in Fig 2A we can then determine the change in $p$ as seen in Eq 4A

$$\delta p = (p' - p) = p_x \left( \frac{R' - R}{x} \right) \tag{4A}$$

Further using the geometry of the system, we can then turn the x component of the momentum into Eq 5A

$$p_x \left( \frac{R' - R}{x} \right) = p \left( \frac{y}{\sqrt{y^2 + r^2}} \right) \left( \frac{R' - R}{x} \right) \tag{5A}$$

Inserting this back into Eq 3A then generates Eq. 7 from the main manuscript.

$$E = \left(\frac{y}{\sqrt{y^2+r^2}}\right)\left(\frac{R'-R}{x}\right)\frac{pc\beta}{\delta xq} \qquad (6A)$$

## Appendix C

Basic simulations of the electron beam radiography experiments were generated using the software G4 Beamline [37]. The simulations are limited to $<10^7$ electrons due to computational limits, which causes the resulting images to be less clear than the actual radiographs due to aliasing. These simulations are also limited by their simple beam geometry and lack of complex electric field gradients. They were only used to guide investigation of field induced magnification changes, rather than confirm particular electric field values for a given experiment. To test the effect of a simple radial electric field, simulations were generated with a W target in the 3.58 cm configuration and radial electric fields ranging from 0 GV/m to 2 GV/m. The change in the image was then measured (Fig 3A) and the effect on the image determined.

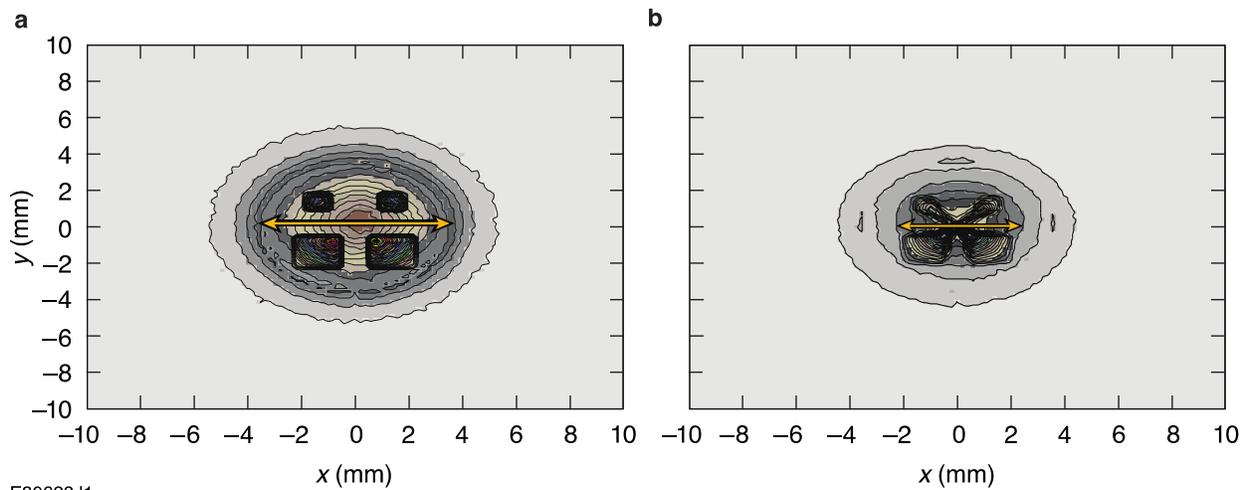

E30628J1

**Figure 3A.** The simulated radiograph of a W test object at 3.58 cm from the source (a) without an electric field and (b) with a 1.5 GV/m radial electric field. The yellow line is representative of the measurement of the target radius in the simulated images and are ~3 mm difference in length.

Magnetic fields, electric fields of opposite sign, electric fields along the thickness of the target and combinations of the options were all tested. It was found that only radial electric fields on the order of 1-3 GV/m matched what was seen in the experiments best (Fig 4A). Magnetic fields both rotated and distorted the image, while lengthwise electric fields provided effectively no change in the image. Combinations of electric and magnetic fields could not prevent visual obvious image distortion and thus, are not considered likely.

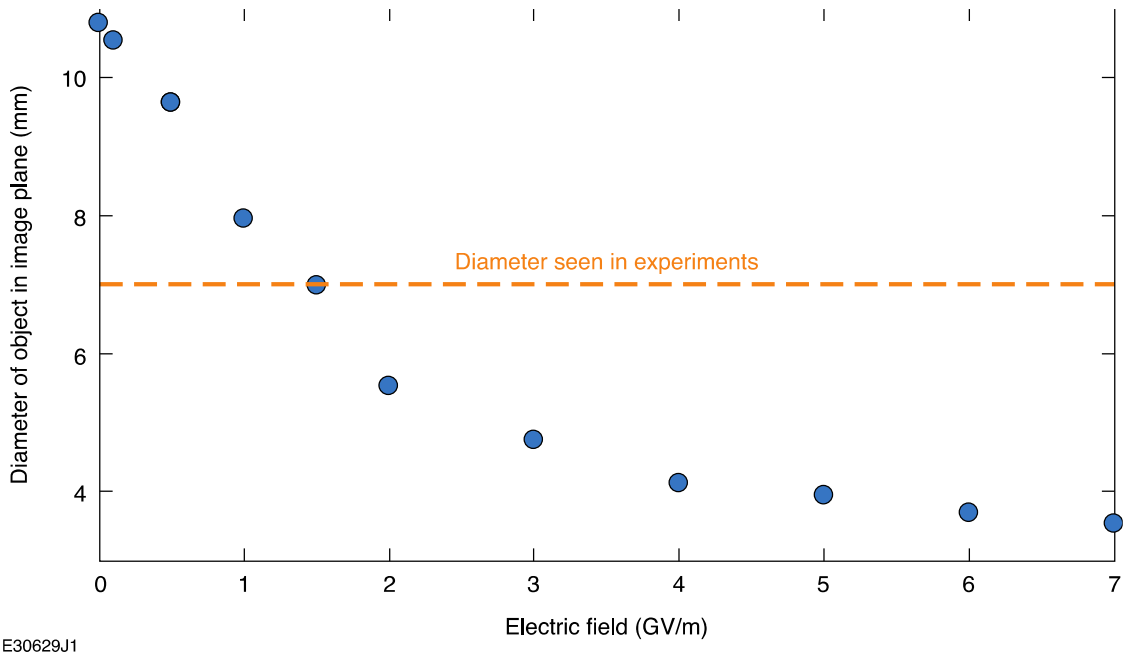

E30629J1

**Figure 4A.** Plot of measured radius of target in G4 Beamline simulation vs radial electric field strength